\documentclass[journal]{IEEEtran}
\usepackage{amsmath}
\usepackage{amsfonts}
\usepackage{amssymb}
\usepackage{graphicx}
\usepackage[cmintegrals]{newtxmath}
\usepackage{float}
\usepackage{cite}

\usepackage{anysize}
\marginsize{1.2cm}{0.85cm}{0.6cm}{1.85cm}

\begin{document}
%
\title{Cooperation in Space: HAPS-Aided Optical Inter-Satellite Connectivity with Opportunistic  Scheduling}


\author{Eylem Erdogan,~\IEEEmembership{Senior Member,~IEEE},~Ibrahim~ Altunbas,~\IEEEmembership{Senior Member,~IEEE},~Gunes Karabulut Kurt,~\IEEEmembership{Senior Member,~IEEE}~and~Halim~Yanikomeroglu~\IEEEmembership{Fellow,~IEEE}

\thanks{E. Erdogan is with the Department of Electrical and Electronics Engineering, Istanbul Medeniyet University,  Istanbul, Turkey, (e-mail: eylem.erdogan@medeniyet.edu.tr). }
\thanks{I. Altunbas is with the Department of Electronics and Communication Engineering, Istanbul Technical University, Istanbul, Turkey, (e-mail: ibraltunbas@itu.edu.tr).}%
\thanks{G.~Karabulut~Kurt is with the Department of Electrical Engineering, Polytechnique Montréal, Montréal, QC, Canada (e-mail: gunes.kurt@polymtl.ca). }%
\thanks{H. Yanikomeroglu is with the Department of Systems and Computer Engineering, Carleton University, Ottawa, ON, Canada, (e-mail: halim@sce.carleton.ca).}

}

\markboth{Journal of \LaTeX\ Class Files,~Vol.~13, No.~9, September~2014}%
{Shell \MakeLowercase{\textit{et al.}}: Bare Demo of IEEEtran.cls for Journals}
%

\maketitle

\begin{abstract}
Issues with tracking, precision pointing, and Doppler shift are the major sources of performance loss in laser inter-satellite communication that can severely decrease the coverage and overall performance of satellite constellations. As a solution to these problems, we propose a cooperation strategy in which a high altitude platform station (HAPS)  staying at a quasi-stationary position contributes to the inter-satellite connectivity. In this setup, the HAPS node uses two different scheduling approaches: one that relies on the zenith angle; the other on instantaneous signal-to-noise ratio. To quantify the performance of the proposed scheme, overall outage probabilities for the two scheduling methods are obtained. In addition, guidelines for the design of practical inter-satellite networks are provided.
\end{abstract}

\begin{IEEEkeywords}
Inter-satellite communications, HAPS-aided transmission, laser satellite communication
\end{IEEEkeywords}

\IEEEpeerreviewmaketitle

\vspace{-0.2cm}
\section{Introduction}

Laser satellite communication (SatCom) using free-space optical communication has attracted considerable interest both from academia and industry in recent years. This technology can provide massive connectivity on the order of thousands of kilometers between various types of satellites, including geostationary-earth-orbit (GEO), medium-earth-orbit (MEO), low-earth-orbit (LEO) satellites, through downlink and uplink communications with ground stations \cite{chaudhry2020free,9324793,9205852,erdogan2020site}. Laser SatCom can also provide numerous advantages compared to its RF counterpart, including lower power consumption, lower cost, and reduced mass \cite{kota2003broadband}, \cite{yahia2021haps}. Furthermore, optical communication is immune to interference, and it does not need regulatory restrictions for the frequency use. Due to these advantages, optical communication has been used in the constellations of SpaceX's Starlink and OneWeb to provide inter-satellite connectivity between a massive number of LEO satellites.

In LEO satellite constellations, laser inter-satellite communication (ISC) with faultless tracking and precision-pointing is of utmost importance, as the satellites move at a speed of $27000$ km/h and complete a full orbit every $90$-$120$ minutes \cite{sodnik2010optical}. This relative motion causes a significantly high Doppler shift at the receiver node and changes the frequency of the received signal. Because of this extremely high relative motion, it is almost impossible to create inter-satellite connectivity between multiple number of satellites. Furthermore, satellite vibrations and various noise sources, including shot noise, thermal noise, and cosmic dust, can cause deviation in the signal and misalignment occurs in laser ISCs \cite{kaushal2016optical}. 
{

\textbf{Prior related research:} }In the literature, various studies have focused on laser ISCs. In \cite{patnaik2012inter}, the authors proposed a new laser ISC model, which could achieve a better bit error rate as long as line-of-sight (LOS) connectivity was established, whereas \cite{liu2017performance} considered the performance degradation of {$\lambda = 1550$ nm wavelength laser} ISC due to ionizing radiation in space. { Furthermore, \cite{tawfik2021performance} investigated the performance of laser ISC between GEO to LEO satellites, where a new ISC system was proposed with  suitable operating wavelength. On the contrary, \cite{song2017impact} examined the effect of pointing errors on the performance of laser ISC, where closed-form average bit error rate was obtained, and \cite{guelman2004acquisition} discussed the acquisition, pointing and tracking concept for the laser ISCs, where a cooperative control was proposed for the satellites.  }

{
\textbf{Motivation and our contribution:} 
}
 In laser ISCs, the above-mentioned Doppler shift, tracking, and precision-pointing errors are major sources of loss, and the current literature has focused on the performance degradation of these impairments. { Furthermore, none of the previous works have considered inter-satellite connectivity between multiple number of satellites.} Different from the current literature, in this paper, we propose a high altitude platform station (HAPS)-aided laser ISC cooperation model. In this model, a HAPS node, defined by the International Telecommunications Union (ITU) as an aerial terminal that can stay at a quasi-stationary position in the stratosphere at an altitude of about $20-50$ km, assists the ISC between multiple number of LEO satellites by using two different scheduling protocols. { In this structure, Doppler shift can be decreased to a certain level that do not deteriorate the communication performance, and precision and tracking problems can be neglected due to the quasi-stationary position of HAPS systems. However, stratospheric attenuations due to volcanic activity, and  stratospheric turbulence related with pressure and temperature changes induced fading should be taken into consideration within satellite and HAPS communications.} In proposing this model, we employ two opportunistic scheduling methods. In the first method, the HAPS node schedules the best satellite which has the lowest zenith angle. In the second method, the satellite with the highest instantaneous signal-to-noise ratio (SNR) is scheduled. The paper makes the following contributions:

\begin{itemize}
	\item { We propose a new model in which a HAPS node can assist the laser ISC when the direct connectivity can not be established between multiple LEO satellites due to high elevation angle, long distance or the curvature of Earth. 
	\item  Even though there are a few number of papers about optical satellite-HAPS communications \cite{R2021HAPSBasedRF,huang2021uplink}, none of them have considered atmospheric layer-based attenuations. Motivated by this gap, this paper accurately models the stratospheric attenuation with the aid of \cite{giggenbach2002stratospheric} and by elaborating the Beer-Lambert law. Furthermore, stratospheric turbulence-induced fading is modeled by using Exponentiated-Weibull fading.} This enables satellite-HAPS communication to be accurately analyzed. 
	\item The paper proposes a novel cooperation strategy where a quasi-stationary HAPS system aids the ISC by using two scheduling methods: 1) scheduling the satellite with the lowest zenith angle; (2) scheduling the satellite with the highest instantaneous SNR. {By doing so, global connectivity can be established for the ISC thanks to the larger footprint of HAPS systems.}
	\item To quantify the performance of the proposed scheme, we derive outage probability for both scheduling methods. Furthermore, we provide guidelines that can be useful in the design of HAPS-aided laser ISCs. 
\end{itemize} 

\begin{figure}[t!]
	\centering
	\includegraphics[width=2.7in]{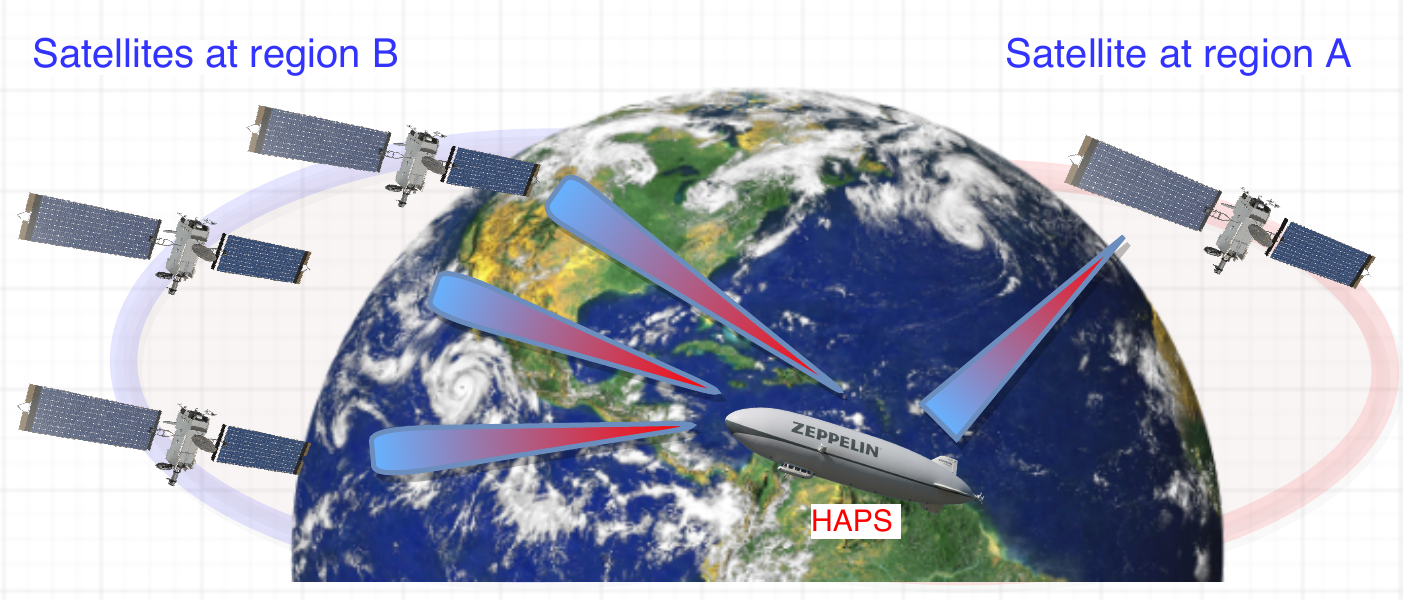}
	\caption{Illustration of the HAPS-aided laser inter-satellite communication model}
	\label{fig_1}
\end{figure}
%

\section{System Model}

Here, we consider a dual-hop laser ISC model where a decode-and-forward (DF) HAPS node assists the transmission to provide seamless connectivity between multiple satellites. In this model, an LEO satellite deployed in a circular orbit at region $A$ seeks to communicate with $N$ number of LEO satellites that are available for communication in region $B$. { The direct connectivity may not be established between those satellites as the maximum elevation angle should be at least $-30$ degrees for two consecutive LEO satellites \cite{carrizo2020optical}. Furthermore, when the satellites are at different regions, the distance can go up to $8000$ km, so that the LOS connectivity cannot be satisfied due to the curvature of Earth.} To overcome these challenges, HAPS terminal ($H$) aids the laser ISC between the satellite at region $A$ and the satellites at region $B$, as shown in Fig. \ref{fig_1}.

For this model, let us assume that $I_{A, H}$ is given as the aggregated channel coefficient between $A$ and $H$, $x_A$ is the satellite information signal, $P_{A}$ shows the satellite's transmit power, and $n_H$ is the additive Gaussian noise with $N_0$ one-sided noise power spectral density. Therefore, the received signal in the first phase of the communication at $H$ can be written as $y_H = \sqrt{P_{A}}I_{A, H}x_{A} +  n_H$.  In the second phase of the communication, $H$ detects the information, schedules the best satellite at region $B$ on the basis of opportunistic scheduling, and forwards the tracking-enabled beam to the scheduled satellite. The received signal at scheduled satellite $K$ can be expressed as $y_{B_K} = \sqrt{P_H}I_{H, B_K}\tilde{x}_{A} +  n_{B}$, where $\tilde{x}_{{A}}$ is the detected  information of satellite at $A$, $I_{H, B_K}$ is the aggregated channel coefficient between $H$ and  $B_K$, and $P_H$ is the transmit power of the HAPS. As $H$ has full decoding capability, DF relaying is adopted to the HAPS system, so that the overall SNR can be expressed as $\gamma_{o} = \min\Big(\gamma_{A , H},\gamma_{H , B_K}\Big)$, where $\gamma_{A,H} = \bar{\gamma}_{A,H}I_{A,H}^2$ and  $\gamma_{H,B_K} = \bar{\gamma}_{H , B_K}I_{H, B_K}^2$ are the equivalent instantaneous SNRs, $\bar{\gamma}_{A,H} = \frac{P_A}{N_0}E[I_{A,H}^2]$ and $\bar{\gamma}_{H,B_K} = \frac{P_H}{N_0}E[I_{H,B_K}^2] $ are the average SNRs, $E[\cdot]$ denotes the expected value, and $E[I_{A,H}^2] = E[I_{H,B_K}^2] = 1$.

\subsection{Aggregated Channel Model}

In the HAPS-aided laser ISC model we propose, the negative effects of Doppler shift, tracking, and pointing can be mitigated thanks to the quasi-stationary position of the HAPS node. {However, stratospheric attenuation and turbulence depending on the non-static stratosphere and mesosphere layers are the major sources of loss. Therefore, by considering both stratospheric attenuation and turbulence-induced fading, we can  express the aggregated channel between $A , H$ and $H , B_K$ as $I_{z} = f_zg_z$  \cite{erdogan2020site}, \cite{yahia2021haps}, where $f_{z}$ denotes the stratospheric turbulence and $g_{z}$ stands for the stratospheric attenuation for $z\in\{A,H ; H,B_K\}$.}

\subsubsection{Stratospheric attenuation}
Stratospheric attenuation can be defined as the attenuation of the optical beam due to aerosols, dust, and smoke in the stratosphere and mesosphere layers. More precisely, there are five major sources of stratospheric attenuation: 1) cosmic dust, radiation-based absorption, and noctilucent clouds consisting of ice crystals in the mesosphere; 2) stratospheric aerosols produced by volcanic activity and its resulting dust; 3) molecular absorption; 4) scattering by ice crystals in extreme weather conditions, like hurricanes when the clouds move up into the stratosphere; 5) polar stratospheric clouds at altitudes of $15000-25000$ m in winter and more in northerly latitudes. Among these sources of attenuation, volcanic activity has the biggest impact on performance loss, since volcanic eruptions can bring a significant amount of sulfuric acid and volcanic ash to the stratosphere \cite{giggenbach2002stratospheric}. 

For the proposed scenario, we assume that stratosphere and mesosphere layers are affective in the HAPS to satellite communication. In this context, if we assume that these layers have two different attenuation parameters and they are affective at different distances. Therefore, the well-known Beer-Lambert law can be rewritten as \cite{erdogan2020site} 
\small
\begin{align}
g_{z} = \exp\Big(-\Big(\Theta^{(1)}_{z}\rho L_{z}+\Theta^{(2)}_{z} (1-\rho)L_{z})\Big),
\label{EQN:1}
\end{align}  
\normalsize
where $0 \leq \rho \leq 1$ shows the fraction of the distance that the beam is affected from the attenuation, $\Theta^{(1)}_{z}$ is the mesospheric attenuation coefficient, $\Theta^{(2)}_{z}$ is the stratospheric attenuation coefficient, $L_{z} = (h_\text{SAT}-h_\text{HAPS})\sec(\zeta_z)$ is the propagation distance, $h_\text{SAT}$ is the satellite altitude, $h_\text{HAPS}$ is the HAPS altitude, and $\zeta_z$ is the zenith angle. In this scheme, the laser beam is affected by the radiation-based absorption in the first $\rho L_z$ distance (mostly in the mesosphere), whereas in the $(1-\rho) L_z$ distance (mostly in the stratosphere) volcanic activity-based ingredients of sulphuric acid and volcanic ash are the predominant source of losses{\footnote{These attenuation coefficients can be obtained from \cite{giggenbach2002stratospheric} by using curve fitting and interpolation techniques.}}.


\subsubsection{Stratospheric turbulence}
Stratospheric turbulence can be generated by the pressure and temperature differences between a satellite and HAPS node, which can result in irradiance fluctuations. Furthermore, non-static high altitude winds can cause scintillation and change the refractive structure index parameter. Stratospheric turbulence can be modeled with EW fading, as it is the best fit for aperture averaging on the basis of empirical results. The cumulative distribution function (cdf) of $f_{z}$ can be expressed as 
\small
\begin{align}
{\mathcal{F}}_{f_{z}}(I) = \Bigg(1-\exp\Bigg[-\Bigg(\frac{I}{\eta_{z}}\Bigg)^{\beta_{z}}\Bigg]\Bigg)^{\alpha_{z}}, 
\label{EQN:2}	
\end{align}
\normalsize
where $\alpha_{z}$ and $\beta_{z}$ are the shape parameters, and $\eta_{z}$ is the scale parameter \cite{porras2013exponentiated}. In the first hop, aperture averaging is adopted at the HAPS terminal to decrease the effects of stratospheric turbulence and to enhance the overall system performance. With aperture averaging, the scintillation is spatially averaged over the aperture and expressed as \cite[eqn. (40)]{erdogan2020site}
\small
\begin{align}
&\sigma_{I_{A, H}}^2 = 8.7\varrho^{{7}/{6}}(h_\text{SAT}-h_\text{HAPS})^{{5}/{6}}\sec^{{11}/{6}}(\zeta_{{A, H}}) \nonumber \\ &\times \Re\Bigg\{ \int_{h_\text{HAPS}}^{h_\text{SAT}} C_{n}^2(\varpi)\Bigg[\Bigg(\frac{\varrho D_G^2}{16L_{A, H}} + i\frac{\varpi-h_\text{HAPS}}{h_\text{SAT}-h_\text{HAPS}}\Bigg)^{\frac{5}{6}}\nonumber\\ &- \Bigg(\frac{\varrho D_G^2}{16L_{A, H}}\Bigg)^{\frac{5}{6}}\Bigg]d\varpi \Bigg\},
\label{EQN:3}
\end{align}
\normalsize
{where $\Re$ denotes the real number,} $\varrho$ is the optical wave number, $D_G$ is the hard aperture diameter, and $C_{n}^2(\varpi)$ is the altitude ($\varpi$) and stratospheric wind speed ($v_{g}$) dependent refractive index structure parameter. The fading parameters $\alpha_{A, H}$, $\beta_{A, H}$, and $\eta_{A, H}$ can be found as given in \cite[eqn. (13)]{erdogan2020site} depending on the $\sigma_{I_{A, H}}^2$. {In the second time slot, we apply two different scheduling approaches which are described in Section III, and the scintillation index for the second hop can be expressed as 
\small
 \begin{align}
& \sigma_{I_{H, B_k}}^2 =2.2\varrho^{\frac{7}{6}}(h_{\text{SAT}}-h_\text{HAPS})^{5/6}\sec^{11/6}\Big(\zeta_{H, B_k}\Big) \nonumber \\ &\times \Re\Bigg\{ \int_{h_0}^H C_{n}^2(\varpi)\Bigg(1-\frac{\varpi-h_\text{HAPS}}{h_{\text{SAT}}-h_\text{HAPS}}\Bigg)^{\frac{5}{6}}\Bigg(\frac{\varpi-h_\text{HAPS}}{h_{\text{SAT}}-h_\text{HAPS}}\Bigg)^{\frac{5}{6}}d\varpi.
\label{EQN:4}
\end{align}
\normalsize
As can be seen from $(\ref{EQN:4})$, the scintillation index is directly related with the zenith angle between $H$ and $B_k$.}



\section{Satellite Scheduling Strategies}
For the sake of completeness, this section proposes and summarizes two new scheduling approaches for the HAPS-aided laser ISC. 

\subsection{Satellite Scheduling Strategy I (SS-I): Minimization of the Zenith angle}
In the satellite-HAPS communications, zenith angle is a dominant factor as it affects the scintillation index which measures the normalized intensity variance caused by atmospheric turbulence. Therefore, in the proposed strategy, we assume that $H$ estimates all the zenith angles between $H$ and $B_k$ and selects the best satellite at region $B$ with the lowest zenith angle. Mathematically, it can be expressed as
\begin{align}
K = \arg\min_{1\leq k\leq N}\Big[\zeta_{H, B_k} \Big],
\label{EQN:5}
\end{align}
where $k$ shows the satellite index. By using (\ref{EQN:5}), the scintillation index for the scheduled satellite $K$ can be expressed as 
 \small
\begin{align}
& \sigma_{I_{H, B_K}}^2  = \min_{1\leq k\leq N} \Big(\sigma_{I_{H, B_k}}^2\Big)\nonumber\\ & =2.2\varrho^{\frac{7}{6}}(h_{\text{SAT}}-h_\text{HAPS})^{5/6}\sec^{11/6}\Big(\min_{1\leq k\leq N}\Big(\zeta_{H, B_k}\Big) \Big) \nonumber \\ &\times \Re\Bigg\{ \int_{h_0}^H C_{n}^2(\varpi)\Bigg(1-\frac{\varpi-h_\text{HAPS}}{h_{\text{SAT}}-h_\text{HAPS}}\Bigg)^{\frac{5}{6}}\Bigg(\frac{\varpi-h_\text{HAPS}}{h_{\text{SAT}}-h_\text{HAPS}}\Bigg)^{\frac{5}{6}}d\varpi. \label{EQN:6}
\end{align}
\normalsize
Depending on the scintillation index given above, the EW fading severity parameters can be obtained numerically as $\alpha_{H, B_K} = \min\limits_{1\leq k\leq N}(\alpha_{H_k})$,  $\beta_{H, B_K} = \max\limits_{1\leq k\leq N}(\beta_{H_k})$ and $\eta_{H, B_K} = \max\limits_{1\leq k\leq N}(\eta_{H_k})$ by using Monte-Carlo simulations.

\subsection{Scheduling Strategy II (SS-II): Maximization of the Instantaneous SNR}

In the second strategy, we apply the well-known opportunistic scheduling idea where the SNR between $H$ and $B_k$ is maximized as the $B_K$ is scheduled among $K$ number of candidates by an exhaustive search. By doing so, the overall outage performance can be maximized. Please note that in SNR based scheduling, $H$ has to have the knowledge of all $I_{A , B_k}$. Mathematically, the scheduled satellite index can be obtained as 
\begin{align}
K = \arg\max_{1\leq k\leq N}\Big[\gamma_{H, B_k}\Big].
	\label{EQN:7}
\end{align}

\section{Performance Analysis}

This section presents the outage performance analysis of the proposed scheduling strategies given in the previous Section. First, we analyze the cdf of $\gamma_z$. Thereafter, we obtain the outage probabilities of the proposed scheduling strategies

\subsection{Statistical Characterization of the SNR}
By using (\ref{EQN:2}), the cdf of $\gamma_z$ can be expressed as \cite{porras2013exponentiated}
\small
\begin{align}
\mathcal{F}_{\gamma_z}(\gamma) =  \Bigg(1-\exp\Bigg[-\Bigg(\frac{\gamma}{\Big(\eta_zg_z\Big)^2\bar{\gamma}_z}\Bigg)^{\beta_z/2}\Bigg]\Bigg)^{\alpha_z}.
\label{EQN:8}
\end{align}
\normalsize
If we apply the Newton's Binomial theorem ($(x+y)^r = \sum_{k=0}^\infty \binom{k}{r}x^{r-k}y^k$), $F_{\gamma_z}(\gamma)$  can be expressed as \cite{erdogan2019joint}
\begin{align}
\mathcal{F}_{\gamma_z}(\gamma) =  \sum_{\rho=0}^{\infty}\binom{\alpha_z}{\rho}(-1)^\rho\exp\Bigg[-\rho\Bigg(\frac{\gamma}{\Big(\eta_zg_z\Big)^2\bar{\gamma}_z}\Bigg)^{\beta_z/2}\Bigg]\Bigg)^{\alpha_z}.
\label{EQN:9}
\end{align} 
{Even though (\ref{EQN:9}) has infinite series, it converges very fast as  the exponential term in (\ref{EQN:8}) satisfies the convergence conditions of Newton's Binomial theorem as $0 < \exp\Big[-\big(\frac{\gamma}{\big(\eta_zg_z\big)^2\bar{\gamma}_z}\Big)^{\beta_z/2}\Big] < 1$.  }
 
  

\subsection{Outage Probability Analysis}
Outage probability can be defined as the probability that the overall SNR falls below a predetermined threshold for acceptable communication quality, and it can be formulated as $P_{\text{out}} = \Pr[\gamma_o\leq\gamma_{th}]$. 

\subsubsection{Scheduling Strategy I}
For the first scheduling strategy, as the zenith angle is minimized to maximize the outage performance, the minimum zenith angle depending on the scintillation index for the scheduled satellite $K$ can be expressed as
\vspace{-0.5cm}
\small
\begin{align}
&\sigma_{I_{H, B_K}}^2 = \min_{1\leq k\leq N}\nonumber \Big(\sigma_{I_{H, B_k}}^2\Big)\\ & =2.2\varrho^{\frac{7}{6}}(h_{\text{SAT}}-h_\text{HAPS})^{5/6}\sec^{11/6}\Big(\min_{1\leq k\leq N}\Big(\zeta_{H, B_k}\Big) \Big) \nonumber \\ &\times \Re\Bigg\{ \int_{h_0}^H C_{n}^2(\varpi)\Bigg(1-\frac{\varpi-h_\text{HAPS}}{h_{\text{SAT}}-h_\text{HAPS}}\Bigg)^{\frac{5}{6}}\Bigg(\frac{\varpi-h_\text{HAPS}}{h_{\text{SAT}}-h_\text{HAPS}}\Bigg)^{\frac{5}{6}}d\varpi.
\label{EQN:10}
\end{align}
\normalsize
Depending on the scintillation index given above, the EW fading severity parameters can be obtained numerically as $\alpha_{H, B_K} = \min\limits_{1\leq k\leq N}(\alpha_{H_k})$,  $\beta_{H, B_K} = \max\limits_{1\leq k\leq N}(\beta_{H_k})$ and $\eta_{H, B_K} = \max\limits_{1\leq k\leq N}(\eta_{H_k})$ by using Monte-Carlo simulations. Therefore, for the first scheduling strategy, the outage probability can be expressed as 
\vspace{-0.5cm}
\begin{align}
P_{\text{out}}^{\text{SS-I}} & = \Pr\Big[\min\Big(\gamma_{A , H},  \gamma_{H, B_K}\Big) \leq \gamma_{th}\Big] \nonumber\\ 
& = 1 - \Big(1-\mathcal{F}_{\gamma_{A , H}}(\gamma_{th})\Big)\Big(1-\mathcal{F}_{\gamma_{H, B_K}}(\gamma_{th})\Big).
\label{EQN:11}
\end{align}
and with the aid of (\ref{EQN:9}), it can be obtained as given in (\ref{EQN:12}).
\small
\begin{figure*}[t!]
\vspace{-0.4cm}
\begin{align}
P_{\text{out}}^{\text{SS-I}}=1 - \sum_{\rho=1}^{\infty}\sum_{\psi=1}^{\infty}     \binom{\alpha_{A, H}}{\rho}\binom{\alpha_{H, B_K}}{\psi}(-1)^{\rho+\psi+2}\exp\Bigg[-\rho\Bigg(\frac{\gamma_{th}}{\Big(g_{A, H}\eta_{A, H}\Big)^2\bar{\gamma}_{A, H}}\Bigg)^{\frac{\beta_{A, H}}{2}} -\psi\Bigg(\frac{\gamma_{th}}{\Big(g_{H, B_K}\eta_{H, B_K}\Big)^2\bar{\gamma}_{H, B_K}}\Bigg)^{\frac{\beta_{H, B_K}}{2}}\Bigg]. 
\label{EQN:12}
\end{align}
\vspace{-0.35cm}
\hrulefill
\end{figure*}
\normalsize

\subsubsection{Scheduling Strategy II}
In the second strategy, the HAPS terminal schedules the satellite with the highest instantaneous SNR to maximize the overall outage performance. Depending on the above-defined scheduling strategy, the outage probability can be expressed as
\vspace{-0.5cm}
\small 
\begin{align}
P_{\text{out}}^{\text{SS-II}} & = \Pr\Bigg[\min\Bigg(\gamma_{A, H}, \max_{1\leq k\leq N} \Big(\gamma_{H , B_k}\Big)\Bigg) \leq \gamma_{th}\Bigg] \nonumber\\ 
& = 1 - \Bigg(1-\mathcal{F}_{\gamma_{A, H}}(\gamma_{th})\Bigg)\Bigg(1-\prod_{k=1}^{N}\mathcal{F}_{\gamma_{H , B_k}}(\gamma_{th})\Bigg).
\label{EQN:13}
\end{align}
\normalsize
{By substituting (\ref{EQN:9}) into (\ref{EQN:13}), with the aid of order statistics, $P_{\text{out}}^{\text{SS-II}}$ can be obtained as can be seen in (\ref{EQN:14}) at the top of the next page.}
\small
\begin{figure*}[!ht]
\vspace{-0.38cm}
\begin{align}
P_{\text{out}}^{\text{SS-II}} & =1 - \sum_{\rho=1}^{\infty}\binom{\alpha_{A, H}}{\rho}(-1)^{\rho+1}\exp\Bigg[-\rho\Bigg(\frac{\gamma_{th}}{\Big(g_{A, H}\eta_{A, H}\Big)^2\bar{\gamma}_{A, H}}\Bigg)^{\frac{\beta_{A, H}}{2}}\Bigg] \nonumber\\ & \times \Bigg(1- \prod_{k=1}^{N}\sum_{\psi=1}^{\infty}{\rho}\binom{\alpha_{H, B_k}}{\psi}(-1)^{\psi+1}\exp\Bigg[-\psi\Bigg(\frac{\gamma_{th}}{\Big(g_{H, B_k}\eta_{H, B_k}\Big)^2\bar{\gamma}_{H, B_k}}\Bigg)^{\frac{\beta_{H, B_k}}{2}}\Bigg]\Bigg). 
\label{EQN:14}
\end{align}
\vspace{-0.5cm}
\hrulefill
\end{figure*}
\normalsize

\small
\begin{table}[t]
   \renewcommand{\arraystretch}{1.1}
   \centering
\caption{List of Parameters and Values} 
\vspace{-0.2cm}
\label{tab1}
\begin{tabular}{|c|c|}
 \hline
 \multicolumn{2}{|c|}{\textbf{Common Parameters}} \\
\hline \hline\textbf{Parameters} & \textbf{Values} \\ 
\hline Number of satellites ($N $) & $5$ \\
\hline Satellite height ($h_\text{SAT} $) &  $500$ km \\
\hline Zenith angle of the first hop ($\zeta_{A, H}$) & $70^\circ$ \\
\hline Wind speed ($v_g$) &  $60$ m/s \\
\hline Wavelength ($\lambda$) & $1550$ nm \\
\hline SNR threshold ($\gamma_{th}$) & $7$ dB \\
\hline Aperture diameter ($D_G$)  & $0.1$ m \\
\hline Fraction of propagation distance ($\rho$) & 0.1 \\
\hline Mesospheric attenuation $\Big(\Theta^{(1)}_{A, H} = \Theta^{(1)}_{H, B_K}\Big)$ &  $1\times 10^{-5} $ \\
\hline  \hline 
 \multicolumn{2}{|c|}{\textbf{Stratospheric Attenuation Parameters }} \\ \hline
 \hline \textbf{Parameters $\Big(\Theta^{(2)}_{A , H} = \Theta^{(2)}_{H , B_K}\Big)$} & \textbf{Values} \\ 
 \hline Extreme volcanic activity  & $4\times 10^{-3}$ \\
\hline  High volcanic activity & $1\times 10^{-3}$\\
\hline  Moderate volcanic activity & $1\times 10^{-4}$ \\
\hline
\end{tabular}
\vspace{-0.4cm}
\end{table}
\normalsize


\begin{figure}[H]
	\centering
	\includegraphics[width=2.5in]{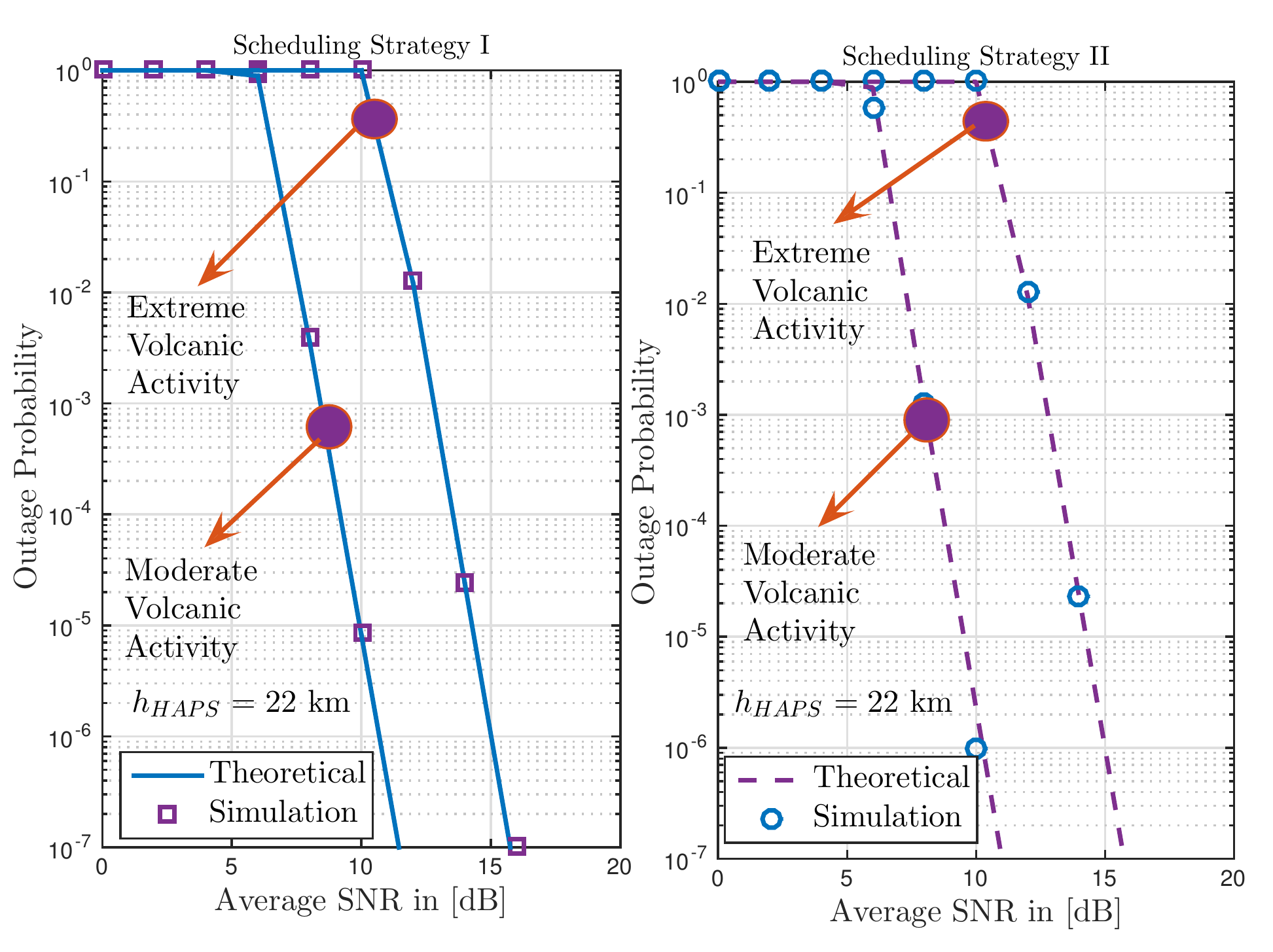}
	\vspace{-0.3cm}
	\caption{Outage probability of the proposed scheduling strategies for varios HAPS altitudes under moderate, high, and extreme volcanic activities.}
	\vspace{-0.5cm}
	\label{fig_2}
\end{figure}
\begin{figure}[H]
	\centering
	\includegraphics[width=2.5in]{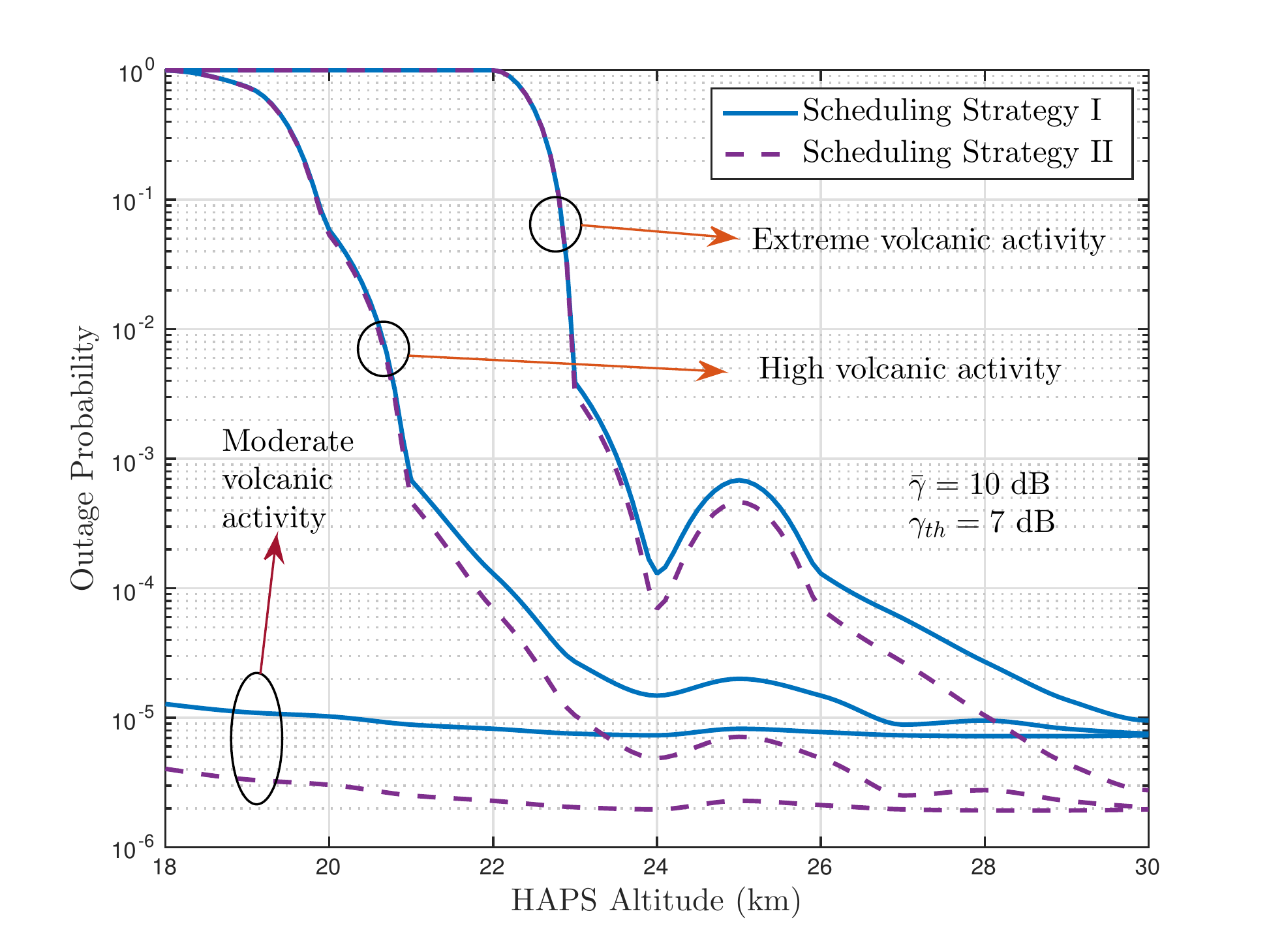}
	\vspace{-0.3cm}
	\caption{Outage probability performance of the proposed scheduling strategies for moderate, high, and extreme volcanic activities.}
	\vspace{-0.44cm}
	\label{fig_3}
\end{figure}

\vspace{-0.3cm}
{
\subsection{Diversity Gain}
 To obtain the diversity gain of SS-I and SS-II, we first express the high SNR approach of (\ref{EQN:11}) and (\ref{EQN:13}) as given in \cite[eqn. (14)]{erdogan2019joint}. Thereafter, by exploiting the asymptotic expansion of $\exp(-x/a) = 1 - x/a$ into (\ref{EQN:8}), than after a few manipulations, the diversity gains can be found as{\footnote{{Please note that the asymptotic analysis can not be presented here in detail due to page limitations.} }}
\small
\begin{align}
& \mathcal{G}_d^\text{SS-I} = \min(\alpha_{A,H},\beta_{A,H_K}) \nonumber\\
& \mathcal{G}_d^\text{SS-II} =  \sum_{k=1}^N\min(\alpha_{A,H},\beta_{A,H_k}).
	\label{EQN:15}
\end{align}
 }
 \normalsize
\vspace{-0.3cm}
\section{Numerical Results}

{In this section, theoretical results are verified by a set of Monte-Carlo simulations. In the simulations, we assume that LEO satellites are deployed on a circular orbit at $500$ km altitude above ground level. Furthermore, $\lambda = 1550$ nm is chosen as the wavelength, the stratospheric wind speed is set to $60$ m/s to demonstrate non-static high altitude winds, average SNRs are assumed as equal ($ \bar{\gamma}_{A , H} = \bar{\gamma}_{H , B_K} = \bar{\gamma} $), HAPS altitude is given as $h_\text{HAPS} = 22$ km unless otherwise stated, the SNR threshold is set to $\gamma_{th} = 7$ dB for acceptable communication quality, and the zenith angles for satellites in region $B$ are randomly set to $81^\circ$, $73^\circ$, $66^\circ$, $77^\circ$, and  $61^\circ$. In addition, $\rho$ is taken as $0.1$ as the first $0.1\times L$ km can be considered as stratosphere and mesosphere layers. Table I presents all simulation parameters for $h_\text{HAPS} = 22$ km based on the empirical parameters given in \cite{giggenbach2002stratospheric}. }


Fig. 2 shows the outage probability performance of the proposed setup as a function of the average SNR for the two different scheduling approaches by considering $h_\text{HAPS} = 22$ km. As we can see from the figure, the theoretical curves, which are shown with solid lines, are in good agreement with the simulations, generated with marker symbols. Furthermore, we observe that the overall outage performance of the proposed scheduling approaches decrease as the density of the volcanic activity increases. 

Fig. 3 compares the two scheduling approaches in terms of HAPS distance for various levels of volcanic activity when the average SNR is set to $\bar{\gamma} = 10$ dB. The figure shows that scheduling Strategy I achieves very similar outage performance to the second strategy for almost all volcanic activity levels up to $22$ km. It is worth noting that above $22$ km, the second scheduling slightly outperforms the first in almost all volcanic activity levels. Furthermore, we can also observe that the outage performance of the proposed scheme worsens between $24$ to $26$ km HAPS altitude due to higher volcanic activity \cite{giggenbach2002stratospheric}.

\subsection{Design Guidelines}
In this section, we provide some important guidelines that can be useful in the design of HAPS-aided laser ISCs. 
\begin{itemize}
\item {The simulations have shown that the proposed cooperation architecture can achieve $10^{-7}$ outage probability performance at about $15$ dB even in the presence of extreme volcanic activity, which can be a sufficient performance for optical ISCs.} 
\item The simulations have shown that a zenith angle-based selection can be preferable in HAPS-aided laser ISC as it is simple and achieves excellent outage performance.
\item  It is important to note that the HAPS altitude can be crucial in HAPS-aided laser ISC, as the stratospheric attenuation level depends on the altitude. 
\item {The proposed cooperation strategy can bring two realistic advantages: 1) The distant LEO satellites can be interconnected with the aid of a quasi-stationary HAPS. 2) As the HAPS system which has a larger footprint towards the space, we can establish global connectivity towards multiple satellites.}

\end{itemize}


\vspace{-0.3cm}
\section{Conclusion}

This paper proposed a cooperation strategy for the satellite communications where a HAPS system aided inter-satellite communication (ISC) by using two different scheduling approaches, one that relied on zenith angle and the other that relied on instantaneous SNR. The outage probability results, verified by simulations, showed that the proposed cooperation method can be a viable solution to the tracking and precision-pointing problems in laser ISCs. Furthermore, we provided some important guidelines that can be useful in the design of practical HAPS-aided laser ISCs.

In the future work, we will elaborate the cooperation idea by including the terrestrial users and unmanned aerial vehicles. Furthermore, we can consider pointing errors, and derive bit error probability and ergodic capacity.

\vspace{-0.3cm}
\bibliographystyle{IEEEtran}

\bibliography{Refs.bib}



\end{document}